\documentclass[pra,twocolumn,showpacs,superscriptaddress]{revtex4-1}
\usepackage{amsmath,amscd,amssymb,color,amsthm}
\usepackage{graphicx,amsfonts,dsfont}
\usepackage{epstopdf}
\usepackage{hyperref}
\usepackage{enumerate,bbold}
\usepackage{array,comment,bm,braket}

\begin{document}
\title{Experimental construction of a symmetric three-qubit
entangled state and its utility in testing the
violation of a  Bell inequality 
on an NMR quantum simulator}
\author{Dileep Singh}
\email{dileepsingh@iisermohali.ac.in}
\affiliation{Department of Physical Sciences, Indian
Institute of Science Education \& 
Research Mohali, Sector 81 SAS Nagar, 
Manauli PO 140306 Punjab, India.}
\author{Vaishali Gulati}
\email{vaishaligulati00@gmail.com}
\affiliation{Department of Physical Sciences, Indian
Institute of Science Education \& 
Research Mohali, Sector 81 SAS Nagar, 
Manauli PO 140306 Punjab, India.}
\author{Arvind}
\email{arvind@iisermohali.ac.in}
\affiliation{Department of Physical Sciences, Indian
Institute of Science Education \& 
Research Mohali, Sector 81 SAS Nagar, 
Manauli PO 140306 Punjab, India.}
\affiliation{Vice Chancellor, Punjabi University Patiala,
147002, Punjab, India}
\author{Kavita Dorai}
\email{kavita@iisermohali.ac.in}
\affiliation{Department of Physical Sciences, Indian
Institute of Science Education \& 
Research Mohali, Sector 81 SAS Nagar, 
Manauli PO 140306 Punjab, India.}
\begin{abstract}
We designed a quantum circuit to prepare a permutation-symmetric maximally
entangled three-qubit state called the $\vert {\rm S} \rangle$ state and
experimentally created it on an NMR quantum processor.  The presence of
entanglement in the state was certified by computing two different entanglement
measures, namely negativity and concurrence.  We used the $\vert {\rm S}
\rangle$ state in conjunction with a set of maximally incompatible local
measurements, to demonstrate the maximal violation of inequality number $26$ in
Sliwa's classification scheme, which is a tight Bell inequality for the (3,2,2)
scenario i.e. the three party, two measurement settings and two measurement
outcomes scenario. 
\end{abstract} 
\maketitle 
\section{Introduction}
\label{sec:intro}
Entanglement is intrinsic to quantum systems and its production,
characterization and protection have been well studied, with a view to using it
as a resource for quantum information processing~\cite{horodecki}.  Three-qubit
entanglement has been extensively characterized and it was showed that three
qubits can be entangled in two inequivalent
ways~\cite{dur-pra-00,sabin-epjd-2008}.  Permutation symmetric entangled states
i.e. states that remain invariant under pair-wise qubit swapping, have been
proven to be useful for a wide variety of quantum information
tasks~\cite{hayashi-pra-2008,toth-prl-2009,mathonet-pra-2010}.  The entanglement
of symmetric multipartite states was characterized using a geometric
entanglement measure~\cite{aulbach-njp-2010,aulbach-ijqi-2012}, and  quantum
circuits were proposed to efficiently construct genuinely entangled
permutation-invariant multipartite quantum states~\cite{burchardt-pra-2021}.  

It is important to design experimental schemes to generate and characterize
different kinds of multipartite entangled states.  In this context, our previous
work on three-qubit entanglement focused on experimentally implementing a
canonical form for three-qubit pure states, from which all other states can be
constructed, including the GHZ and W entangled states~\cite{dogra-pra-2015-1}.
Interestingly, we were able to show that for nearly all pure states (except
those belonging to the GHZ class), the two-qubit reduced states were sufficient
to reconstruct the whole three-qubit state.  We experimentally constructed a
novel three-qubit entangled W$\bar{\rm W}$ state, which is interconvertible with
the GHZ state, but stores information about multipartite correlations in a
totally different way~\cite{dogra-pra-2015-2}.  While studying the decoherence
of entangled states, we found that the GHZ-class of states are the most fragile,
while the W-class of states are the most robust against the natural noise
present in the NMR system~\cite{singh-pra-18}.

Nonlocal correlations in a spatially separated multipartite quantum system have
an inherent measurement statistics which is not classically
reproducible~\cite{bell-p-64} and their presence can be revealed via the
violation of Bell-type inequalities such as the Clauser-Horne-Shimony-Holt
(CHSH) inequality~\cite{peres-fp-99}.  Nonlocality has potential applications in
secure communication~\cite{buhrman-rmp-10} and quantum
information~\cite{chen-pra-02}.  $N$-partite generalization of Wigner's argument
was used to obtain a Bell-type inequality and its violation by GHZ, Cluster and
W states was investigated~\cite{home-pra-2015}.  Nonlocality of three-qubit pure
states has been explored by looking at the violation of the CHSH
inequality~\cite{ozeki-ijqi-20,anjali-qip-21,tendick-prr-22}.  Entanglement has
to be present in the state for nonlocal correlations to
exist~\cite{liang-pra-11},  and several studies have focused on understanding
the connection between nonlocality and
entanglement~\cite{werner-pra-89,acin-pra-06,cavalcanti-prl-16}.
Incompatibility is another important property of quantum systems which states
that the outcome of certain observables cannot be obtained
simultaneously~\cite{heinosaari-pra-15}.  Incompatibility of measurements is
necessary for the violation of any Bell
inequality~\cite{uola-prl-14,quintino-pra-16,bene-njp-18}.  It was demonstrated
that measurement compatibility can be used to both detect Bell nonlocality and
to certify entanglement in a device-independent
manner~\cite{temistocles-pra-19}.  The question of how incompatible the
measurements have to be in order to achieve a violation of a Bell inequality was
explored and quantifiers of measurement incompatibility were
devised~\cite{chen-prr-21}.  The (3,2,2) scenario i.e. the three party, two
measurement settings per party and two measurement outcomes scenario provides
detailed information about the combination of entanglement and incompatibility
that could lead to maximal nonlocality for tight Bell
inequalities~\cite{rosa-pra-16}.  Only Bell, GHZ and $\vert {\rm S}
\rangle$ states produce maximal violation of a tight Bell inequality via
maximally incompatible measurements~\cite{mermin-prl-90,acin-pra-02}.  The
experimental demonstration of the maximum quantum violation of tight Bell
inequalities with GHZ and Bell and the $\vert {\rm S} \rangle$ states has been
demonstrated using a photonic
set-up~\cite{erven-nat-2014,poh-prl-15,anwer-pra-19} as well with ion
traps~\cite{lanyon-prl-14}.

In this work, we designed and experimentally implemented a quantum circuit to
prepare the three-qubit symmetric entangled $\vert {\rm S} \rangle$ state on an
NMR quantum processor.  We certified the entanglement of the $\vert {\rm S}
\rangle$ state by calculating the experimental values of the tripartite
negativity and the concurrence.  We then demonstrated the utility of the $\vert
{\rm S} \rangle$ state by experimentally simulating the violation of the tight
Bell inequality (number $26$ in Sliwa's classification scheme) in the $(3, 2,
2)$ scenario~\cite{rosa-pra-16}.  Maximal incompatibility was checked by adding
compatible measurements in the existing measurement setting.  We refer to our
experiment as a ``simulation'', since in an NMR set-up, the qubits are realized
by nuclear spins bound in the same molecule and separated by a few angstroms, so
there is no actual spatial separation or nonlocality~\cite{dileep-jmro-22}.

The material in this paper is arranged as follows: Section~\ref{sec:theo1}
briefly describes the theoretical framework of the entanglement properties of
the $\vert {\rm S} \rangle$ state and the maximal quantum violation of the tight
Bell inequality using the entangled $\vert S \rangle$ state.
Section~\ref{sec:exp2} describes the experimental construction of the entangled
$\vert {\rm S} \rangle$ state and the simulation of the violation of the tight
Bell inequality, inequality number $26$ in Sliwa's classification in the
$(3,2,2)$ scenario, on a three-qubit NMR quantum computer.  The NMR experimental
details are summarized in Section~\ref{nmr-expt} and the details of the quantum
circuit and NMR pulse sequence to construct the $\vert S \rangle$ state are
given in Section~\ref{s-expt}.  The experimental certification of the presence
of entanglement in the $\vert {\rm S} \rangle$ state is given in
Section~\ref{entang-expt}.  The results of the experimental violation of the
inequality  are given in Section~\ref{ineq-expt}.  Section~\ref{sec:con3}
contains a few concluding remarks.

\section{Theoretical background} 
\label{sec:theo1}  
Three-qubit entanglement with genuine three-party
entanglement, has been well studied and falls into one of the
two inequivalent classes namely, the GHZ  class and the W
class~\cite{dur-pra-00}. 
The $\vert
{\rm S} \rangle$ state which is a three-qubit permutation-symmetric
entangled state, has been defined as~\cite{rosa-pra-16}:
\begin{equation} 
\vert {\rm S} \rangle = \frac{1}{\sqrt{6}} 
(\vert 001 \rangle + \vert 010 \rangle -
\vert 100 \rangle) + \frac{1}{\sqrt{2}} \vert 111 \rangle 
\label{eq:state_1}
\end{equation}
The $\vert {\rm S} \rangle$ belongs to the W class of states
and therefore is inequivalent to the GHZ state under 
local operations and classical communication (LOCC).
The qubit-qubit concurrences for this state are all equal to
0.244, as opposed to  the W state whose qubit-qubit concurrences
are all equal to 0.667~\cite{anwer-pra-19}.  The $\vert {\rm S} \rangle$ 
state has
interesting properties and as we shall see, plays an important role in the
context of violation of Bell's inequalities for the (3,2,2)
scenario.

Quantum nonlocality has been classified according to the
combinations of entangled states and incompatible
measurements for 46 classes of tight Bell inequalities in
the (3,2,2) scenario~\cite{rosa-pra-16}.  The inequality
number 26 in Sliwa's classification scheme is given by:
\begin{equation}
\begin{aligned}
T_{26} &=  \langle A_0 \rangle + \langle B_0 \rangle +
\langle A_0 B_0 \rangle  + 2 \langle A_1 B_1 \rangle +
\langle C_0 \rangle \\ &  + \langle A_0 C_0  \rangle
+\langle B_0 C_0 \rangle - \langle A_0 B_0 C_0 \rangle - 2
\langle A_1 B_1 C_0 \rangle  \\ & +  2 \langle A_1 C_1
\rangle - 2 \langle A_1 B_0 C_1 \rangle - 2 \langle  B_1 C_1
\rangle \\ & + \langle A_0 B_1 C_1 \rangle  \leq 5
\end{aligned}
\label{eq:ineq_1}  
\end{equation}
where $A_i$, $B_i$, $C_i$ are the measurement settings of
the first, second and
third party, respectively. Each observable 
is dichotomous and can have $\pm 1$ outcomes.  
The maximum quantum violation of the inequality number 26 in
Sliwa's classification scheme is achieved using 
the entangled $\vert {\rm S} \rangle$ state
given in Eq.~\ref{eq:state_1}~\cite{siliva-pla-03}. 

For the particular set of observables $A_0 = B_0 = C_0 =
\sigma_z$ and $A_1 = B_1 = C_1 = \sigma_x$ which are
maximally incompatible according to any quantifier of
incompatibility, the $\vert {\rm S} \rangle$ state produces
the maximum quantum violation given by~\cite{rosa-pra-16}:
\begin{equation} T_{26} = 1 + 4\sqrt{3} \approx 7.928
\label{eq:violation}  
\end{equation}
\section{Experimental construction of $\vert {\rm S} \rangle$
state and violation of Bell inequality}
\label{sec:exp2}  
\subsection{NMR experimental details}
\label{nmr-expt}
We used
the molecule iodotrifluoroethylene dissolved in acetone-d6 with the three
($^{19} \text{F}$) spins encoding the three NMR qubits.
The molecular structure and  other experimental details
are given in Ref.~\cite{singh-pra-18}. 
The three-qubit  Hamiltonian in a rotating
frame is given by: 
\begin{equation}
H = - \sum_{i=1}^{3} (\omega_i - \omega_{RF})I_{iz} + \sum_{i> j, j=1}^{3} 2 \pi
J_{ij}I_{iz}I_{jz} 
\label{eq:hamiltonian}  
\end{equation}
where $I_{iz}$, $\omega_i$, $ J_{ij}$ denote the spin angular momentum operator,
the Larmor frequencies, and the scalar coupling constants, respectively. The
first term in the Hamiltonian represents the Zeeman interaction between the
spins and the applied static magnetic field, while the second term represents
the interaction term. More details are provided in Ref~\cite{oliveira-book-07}. 

The NMR experiments were performed at room temperature and the system was
initialized in a pseudopure (PPS) state, which mimics a pure
state~\cite{cory-pd-98,oliveira-book-07}.  The PPS state $\vert 000 \rangle$ was
achieved via the spatial averaging 
technique~\cite{amandeep-epjd-20} with the
density operator being given by:
\begin{equation}
\rho_{000} = \frac{1 - \epsilon}{8} I_8 + \epsilon \vert 000 \rangle \langle 000 \vert
\label{eq:density operator}  
\end{equation}
where $I_8$ is the $8 \times 8$ identity operator and $\epsilon$ is
proportional to the spin polarization which is $\approx 10^{-5}$
at room temperature.

The NMR pulse sequence for preparing the PPS state $\vert 000 \rangle$ can be
found in Ref.~\cite{singh-pra-18} which gives details of the specific sequence
of RF pulses, four Z gradient pulses, and three time evolution periods that are
used.  We used the Gradient Ascent Pulse Engineering (GRAPE)
technique~\cite{tosner-jmr-09,Schulte-trsa-2012} for the optimization of all the
RF pulses used to construct the PPS state. The GRAPE optimized RF pulses are
robust against RF inhomogeneity, with an average fidelity of $\geq 0.999$. We
used four GRAPE pulses ($U_{P1}, U_{P2}, U_{P3}, U_{P4} $) optimized to reach
the PPS state from the thermal state, where some RF pulses were combined into a
single pulse. The duration of pulses $U_{P1}, U_{P2}, U_{P3}, U_{P4}$ are $500
\mu s$, $9000 \mu s$, $7500 \mu s$, $4000 \mu s$ respectively. The experimental
state was reconstructed  using a least squares constrained convex optimization
technique~\cite{gaikwad-qip-22}.  The set of tomographic operations $\{III, IIY,
IYY, YII, XYX, XXY, XXX \}$ where the $X(Y)$ denotes the single spin operator
implemented by a spin-selective $\frac{\pi}{2}$ pulse and $I$ denotes no
operation were performed to reconstruct the final density operator.  All the
seven tomography spin selective pulses were optimized using GRAPE with the
length of each pulse being $\approx 500 \mu s$. The PPS state $\vert 000
\rangle$ had an experimental state fidelity of  $0.997$. 

\begin{figure}[ht]
\centering
\includegraphics[scale=1.0]{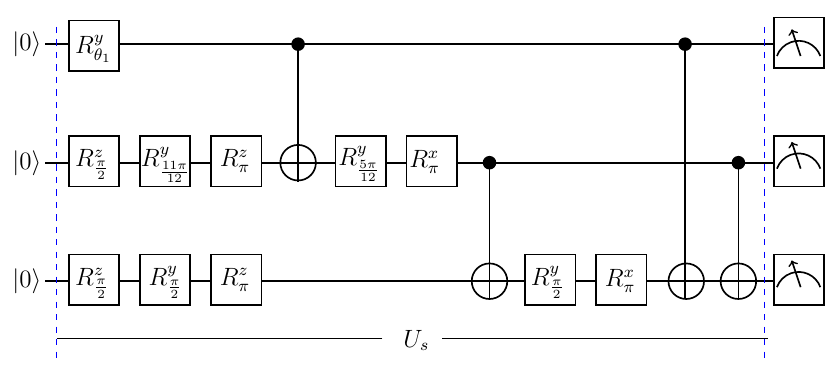}
\caption{Quantum circuit showing the sequence of implementation of the
single-qubit local rotation gates and 
two-qubit controlled-NOT gates required to
construct the $\vert S \rangle$ state 
from the PPS state $\vert 000 \rangle$ state. $U_s$ denotes the 
complete unitary including all quantum gates used for the state preparation.}
\label{nmr_qunt_ckt}
\end{figure}

\begin{figure}[ht]
\centering
\includegraphics[scale=1.0]{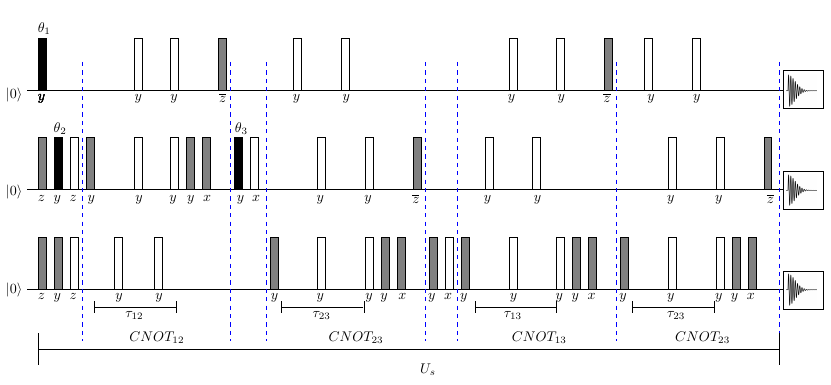}
\caption{NMR pulse sequence corresponding the quantum circuit used to prepare
the $\vert S \rangle$ state 
starting from the PPS state. The black rectangles denote
RF pulse rotations with $\theta_1 = \frac{1.216 \pi}{2}$,
$\theta_2 = \frac{11\pi}{12}$ and   $\theta_2 = \frac{5\pi}{12}$. Grey
rectangles denote $\frac{\pi}{2}$ rotations and unfilled
rectangles denote $\pi$ rotations.  The phase of each pulse is
written below each rectangle. 
The time intervals are set to $\tau_{12} = \frac{1}{2
J_{12}}$, $\tau_{13} = \frac{1}{2 J_{13}}$, $\tau_{23} = \frac{1}{2 J_{23}}$
with $J_{12} = 69.65$ Hz, $J_{13} = 47.67$ Hz, $J_{23} = -128.23$ Hz. 
The blue dotted lines separate the pulses achieving each CNOT gate
and the total combination of all the gates is
denoted by a single unitary $U_s$.}
\label{nmr_qunt_pulse}
\end{figure} 

\subsection{Experimental construction of the $\vert S \rangle$ state}
\label{s-expt} 
After preparing the PPS state, we turn to the experimental preparation of the
tripartite $|S\rangle$ state on the three-qubit NMR system.  The quantum circuit
to prepare the $\vert S\rangle$ state starting from the PPS $\vert 000 \rangle$
state.  is given in Fig.~\ref{nmr_qunt_ckt}. The quantum circuit contains
several one-qubit gates and two-qubit gates. The NMR pulse sequence
corresponding to the quantum circuit is shown in Fig.~\ref{nmr_qunt_pulse}. All
the NMR pulses were numerically optimized using the GRAPE algorithm and we were
able to achieve  high gate fidelities with  relatively small RF pulse durations.
The unitary operator for the entire preparation sequence of
Fig.~\ref{nmr_qunt_pulse} contains four CNOT gates and eleven non-selective
rotations; the entire unitary was generated by a specially crafted single GRAPE
pulse $U_s$ having a duration of $\approx 4600 \mu s$.

\begin{figure}
\centering
\includegraphics[scale=1.0]{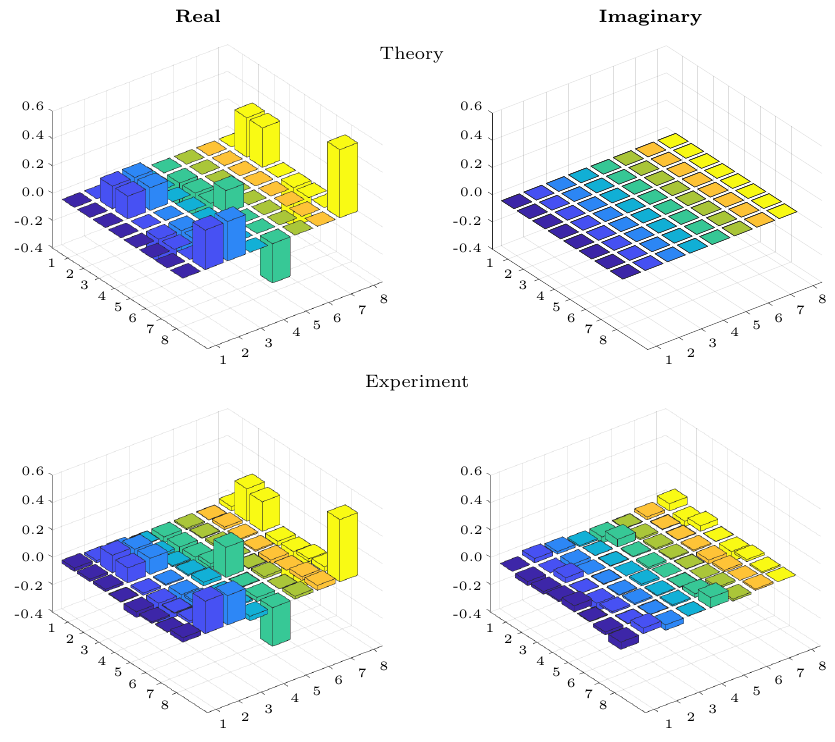}
\caption{The real (left) and imaginary (right) parts of the theoretical and
experimental tomographs of the $\vert S \rangle$ state. 
The labels $1, 2, 3..., 8 $
represent rows and columns encoding the computational basis in binary order
from $\vert 000 \rangle$ to $\vert 111 \rangle$.}
\label{nmr_tomo}
\end{figure}

To check the quality of the prepared  $\vert S\rangle$ state, 
state tomography was
performed using the least squares optimization technique 
and an experimental
state fidelity of $0.949 \pm 0.003$ was
obtained. The tomographs of the 
$\vert S \rangle \langle S \vert $ state are shown in Fig.~\ref{nmr_tomo}. 

\subsection{Entanglement verification of the $\vert S \rangle$ state}
\label{entang-expt}
Next, we experimentally verified the entanglement of the  $\vert S \rangle$
state. To quantify entanglement, we used the well-known tripartite negativity
measure ~\cite{vidal-pra-02}. The tripartite negativity can be calculated using
the bipartite negativity, where bipartite negativity is the absolute value of
the sum of the negative eigenvalues of $\rho^{T_A}$ with $T_A$ denoting the
partial transpose of $\rho$ with respect to the subsystem $A$ in the bipartition
$A|BC$. Negativity is zero if the partial transpose $\rho^{T_A}$ has no negative
eigenvalues. Tripartite negativity $N$ then becomes
$(N_{A|BC}N_{B|AC}N_{C|BA})^\frac{1}{3}$. It ranges from $0$ for a separable
state to $1$ a the maximally entangled state. We computed the tripartite
negativity for the experimentally prepared $\vert S \rangle$ state and obtained
$ N=0.794 \pm 0.015$, which is close to the theoretically predicted value of
$0.943$. This clearly shows the presence of tripartite entanglement in the
$\vert S \rangle$ state. 

We also verified the entanglement of the $\vert S \rangle$ state by using
another entanglement measure termed qubit-qubit
concurrence~\cite{wootters-prl-98}.  The concurrence is calculated from the
eigenvalues $\lambda_1 \geq \lambda_2 \geq \lambda_3 \geq \lambda_4$ of the
matrix $R = \rho (\sigma_y \otimes \sigma_y){\rho}^\ast (\sigma_y \otimes
\sigma_y)$ where $\sigma_y$ is  the Pauli matrix and $\rho$ and ${\rho}^\ast$
denote the density matrix and its complex conjugate, respectively.  The
qubit-qubit concurrence is given by C($\rho$) = max($0, \sqrt{\lambda_1}
- \sqrt{\lambda_2} - \sqrt{\lambda_3} - \sqrt{\lambda_4}$).  The bipartite
  concurrence ranges between $0$ and $1$, with $0$ implying no entanglement is
present in the system and and $1$ indicating maximal entanglement.  The
experimental values we obtained for the qubit-qubit concurrence, range from
$\approx 0.094-0.32$, which verifies the presence of entanglement in the $\vert
S \rangle$ state. 
 
\subsection{Inequality violation}
\label{ineq-expt}
The experimentally prepared $\vert S \rangle$ state was used to test the
inequality $T_{26}$ defined in Eqn~\ref{eq:ineq_1}. The numerical value of
experimental violation we obtain is:
\begin{equation}
T_{26} = 6.531 \pm 0.125
\label{eq:ine_violation}  
\end{equation}
where the observables $A_0 = B_0 = C_0 = \sigma_z$ and $A_1 = B_1 = C_1 = \sigma_x$  are maximally incompatible for each of the three parties.

In addition, we checked the maximal incompatibility of the observables of the
three parties by changing one of the observables out of the six local
measurements while leaving the other five in their initial configuration.  These
measurement modifications led to a decrease in the inequality value of $T_{26}$,
fulfilling the requirement of maximal incompatibility in the three parties for
the $T_{26}$ scenario.
 
\section{Conclusions} 
\label{sec:con3}  
We experimentally prepared the genuinely entangled three-qubit state $\vert {\rm
S} \rangle$ on an NMR quantum processor, and certified its entanglement using
two different entanglement measures namely, negativity, and qubit-qubit
concurrence.  We used the $\vert {\rm S} \rangle$ state to experimentally
simulate the maximum quantum violation of a tripartite tight Bell inequality.
Our results show a clear violation of the tight Bell inequality, revealing the
maximally non-local behavior. We also tested the maximal incompatibility of the
observables by modifying one of the six observables.  
An NMR quantum processor is a good testbed to perform tests of foundational
issues in quantum mechanics. Our results are a step forward in the direction of
unearthing deeper connections between  nonlocality, entanglement, and
incompatibility.
     
\begin{acknowledgments} All the experiments were performed on a Bruker
Avance-III 400 MHz FT-NMR spectrometer at the NMR Research Facility of IISER
Mohali.  A. acknowledges financial support from DST/ICPS/QuST/Theme-1/2019/Q-68.
K~.D. acknowledges financial support from DST/ICPS/QuST/Theme-2/2019/Q-74.
\end{acknowledgments}


%
\end{document}